\documentclass[conference,a4paper]{IEEEtran}
\IEEEoverridecommandlockouts
\usepackage{cite}
\usepackage{amsmath,amssymb,amsfonts}
\usepackage{algorithmic}
\usepackage{graphicx}
\usepackage{textcomp}
\usepackage{xcolor}
\usepackage{boldline} 

\usepackage[colorinlistoftodos, textwidth=3cm, shadow]{todonotes}

\def\BibTeX{{\rm B\kern-.05em{\sc i\kern-.025em b}\kern-.08em
    T\kern-.1667em\lower.7ex\hbox{E}\kern-.125emX}}
\begin{document}

\title{Training Strategies for Autoencoder-based\\ Detection of False Data Injection Attacks
}

\author
{\IEEEauthorblockN{Chenguang~Wang,  Kaikai~Pan, Simon~Tindemans, Peter~Palensky}
\IEEEauthorblockA{\textit{Department of Electrical Sustainable Engineering} \\
\textit{Delft University of Technology}\\
Delft, The Netherlands \\
\{c.wang-8,  k.pan,  s.h.tindemans, p.palensky\}@tudelft.nl}
\thanks{This work is supported by the Chinese Scholarship Council.}}

\allowdisplaybreaks

\IEEEoverridecommandlockouts

\IEEEpubid{\parbox{\columnwidth}{\copyright 2022 IEEE. Personal use of this material is permitted. Permission from IEEE must be obtained for all other uses, in any current or future media, including reprinting/republishing this material for advertising or promotional purposes, creating new collective works, for resale or redistribution to servers or lists, or reuse of any copyrighted component of this work in other works.}\hspace{\columnsep}\makebox[\columnwidth]{ }}

\maketitle

\IEEEpubidadjcol

\begin{abstract}
The security of energy supply in a power grid critically depends on the ability to accurately estimate the state of the system. However, manipulated power flow measurements can potentially hide overloads and bypass the bad data detection scheme to interfere the validity of estimated states. In this paper, we use an autoencoder neural network to detect anomalous system states and investigate the impact of  hyperparameters on the detection performance for false data injection attacks that target power flows. Experimental results on the IEEE 118 bus system indicate that the proposed mechanism has the ability to achieve satisfactory learning efficiency and detection accuracy.

\end{abstract}


\section{Introduction}
State Estimation (SE) is a key element of modern energy management systems (EMS), and an example of the dependency between the physical power system and the ICT infrastructures. It provides the operator with an estimate of the system state, based on power flow measurements delivered by the SCADA system. The estimate of the state guides operational decisions, thus highlighting the importance of ensuring the accuracy and security of SE. However, the SCADA system is vulnerable to a large number of security threats \cite{ChenAbu2011,Giani2009}. As noted by \cite{liu2011false}, false data injection attacks (FDIAs), as a typical class of data integrity attack, can pass the bad data detection (BDD) mechanism within the SE to stay stealthy from the operators. In this light, it is of utmost importance to detect such attacks and respond accordingly\cite{liu2015analyzing,jia2013impact}.

Techniques have been designed to detect stealthy FDIAs in the SE process. Statistical methods, such as an online detection method that leverages load forecasts and generation schedules was described in \cite{Ashok2018}.
In \cite{Pan2019}, the authors introduced a model-based approach that considered the impact of attacks on the dynamics of system trajectories. Notably, such model-based methods can only be accurate if the system dynamics are sufficiently well understood modelled. In practice, developing accurate models that take into account all nonlinearities and uncertainties is difficult or even infeasible,  especially in the context of complex power systems.

In view of this, efforts have been made to address detection challenges by implementing machine learning algorithms\cite{he2017real, james2018online}. Detectors are typically trained in a supervised manner using examples of attacks. However, such examples will be rare (or absent) in real data sets. Moreover, although training data can be enriched with artificially generated attacks, the resulting detector may not be able to recognize potential new attacks by a creative adversary.  \cite{niu2019dynamic} proposed an alternative detection mechanism by dynamically comparing predictions and measurements. Its performance therefore depends on the ability to accurately predict `normal' operating conditions ahead of time. In \cite{Chenguang2020}, we have suggested the use of an anomaly detection (one-class classification) approach to detect attacks, by means of an autoencoder neural network. This has the benefit that no data on attacks is required.

This paper extends our initial work in the following ways:

\begin{itemize}

\item[1)] We describe an autoencoder-based detection approach for FDIAs and investigate the influence of the hyperparameters selection to the training and FDIA detection performance of the proposed mechanism. Experimental results show that the mechanism has the ability to achieve good learning efficiency and detection accuracy.

\item[2)] We use a mixed integer linear program (MILP) reformulation to optimise the number of measurements to be attacked in coordination and evaluate the performance of autoencoder-based detector on these power flow-targeted FDIAs.

\end{itemize}

\section{State Estimation and Data Attacks} 

\subsection{State estimation} 

Considering the power system in steady-state (power flow model), the data collected by sensors in the SCADA network includes line power flow and bus power injection measurements. These measurements, denoted by $z \in \mathbb{R}^{n_z}$, are used to estimate the system state $x \in \mathbb{R}^{n_x}$. For the analysis of cyber-security in SE, it is customary to describe the dependencies of the power flow measurements and the system state through an appropriate linear model, i.e., DC power flow model \cite{TeixeiraSou2015}. In the simplified DC power flow calculations, the measurement vector $z$ refers to active power flow and injection measurements, and the state vector $x$ represents power injections only\footnote{The use of injections is functionally equivalent to the more commonly used phase angle vector $\theta$, but results in more elegant generation and detection of FDIAs.}, and the linear relation can be expressed as
	\begin{equation}
	\label{eq: DC_SE}
	{z} = {H}{x} + {e},
	\end{equation}
where the matrix $H \in \mathbb{R}^{n_z \times n_x}$ describes the dependencies between the measurements and the state, containing the system model information such as the topology of the power network, the transmission line parameters and the placement of the sensors\cite{gonzalez2014powerfactory}. Here ${e} \sim \mathcal{N}(0, {R})$ denotes the measurement noise vector of independent zero-mean Gaussian variables with the covariance matrix ${R} = \mbox{diag}(\sigma_{1}^{2}\, , \, \ldots \, ,\sigma_{n_{z}}^{2})$. In the traditional weighted least squares estimation\cite{sandberg2010security}, an estimated state can be derived through the following
\begin{equation}
\label{eq: WLS_DCSE} 
{\hat{x}} := \mbox{arg} \min\limits_{{x}} ({z} - {H}{x})^{\top} {R}^{-1}({z} - {H}{x}),
\end{equation}
which can be solved as ${\hat{x}} = ({H}^{\top}{R}^{-1}{H})^{-1}{H}^{\top}{R}^{-1}{z}$. 

\subsection{Bad data detection and stealth FDIAs} 

Following the SE process above, the typical BDD mechanism is then conducted to detect erroneous measurements whose statistical properties exceed the presumed standard deviation or mean. To do that, it is customary to define a residual signal, ${r} = {z} - {\hat{z}}$ where $\hat{z}$ is the estimated measurements that satisfy $\hat{z}:=H{\hat{x}}$. Thus the residual signal can be further described by
\begin{equation}
\label{eq: residual}
{r} =({I} - H({H}^{\top}{R}^{-1}{H})^{-1}{H}^{\top}{R}^{-1}){z}.
\end{equation}

Note that an introduced quadratic cost function $J({\hat{x}}) := \lVert {R}^{-1/2}{r} \rVert_{2}^{2}$ follows a generalized chi-squared distribution \cite{pan2018cyber}. With this statistical property, the BDD mechanism uses the hypothesis
test to see whether $J({\hat{x}})$ is larger than expected or not. The detection scheme based on \eqref{eq: residual} has shown a good effectiveness in detecting erroneous data and basic attacks. 
However, prior work \cite{liu2011false} has pointed out that an attack vector $a \in  \mathbb{R}^{n_{z}}$ (resulting in measurements ${z}_{a} = {z} + a$) lies in the range space of the matrix $H$, then the residual signal remains unchanged. We can define $a = H c$ where $c \in \mathbb{R}^{n_{x}}$ is the injected bias. The measurements after such FDIAs are fully consistent with the power system in state $x + c$, so that the BDD method cannot detect such changes.

\section{FDIA DETECTION MECHANISM} 

In this section, we describe how an autoencoder neural network can be used to detect FDIAs. 

\subsection{Autoencoder-based FDIA detector} 

\begin{figure}[htp]
	\centering
	\includegraphics[scale=0.728]{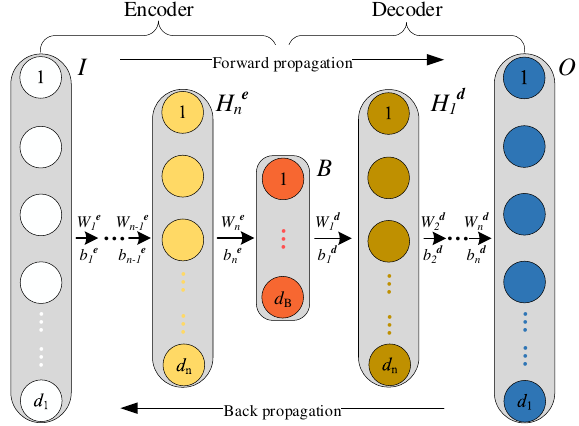}
	\caption{The schematic of the autoencoder.}
	\label{fig:autoencoder}
\end{figure}

Autoencoder neural networks are designed to replicate the original input on the output side with minimal reconstruction errors in an unsupervised manner \cite{sakurada2014anomaly, zhou2017anomaly}. The schematic of the autoencoder algorithm is shown in Fig.~\ref{fig:autoencoder}.  The $d_1$-dimensional data in the input layer $I$ are compressed by the encoder through n hidden layers $H$ to the bottleneck layer $B$. The latent vector with lower dimensionality in the bottleneck layer is then decompressed to output layer $O$.  Weight matrices $W$ and bias vectors $b$ are used in the encoding and decoding process, which can be expressed as 
\begin{subequations}
	\begin{align}\label{eq:y}
	y= \sigma(W_{n}^{e}(\ldots \sigma(W_{1}^{e}z + b_{1}^{e})\ldots) + b_{n}^{e}) \, , 
	\end{align}
	\begin{align}\label{eq:hatz}
	\tilde{z} = \sigma(W_{n}^{d}(\ldots \sigma(W_{1}^{d}y+ b_{1}^{d})\ldots) + b_{n}^{d}) \, , 
	\end{align}
\begin{equation}
\label{eq:Re_error}
{ \tilde{r} }(z) = \lVert {z}-\tilde{z}\rVert^{2}/d_1.
\end{equation}
\end{subequations}
where  $z$ denotes  the input data vector, $y$ represents  the data in the bottleneck layer,  vector $\tilde{z}$ stands for the output data, and $\sigma$ refers to a nonlinear element-wise activation function. The corresponding reconstruction error $\tilde{r}(z)$ is calculated in \eqref{eq:Re_error} as the mean squared reconstruction error.

The main process of using the autoencoder algorithm to detect FDIAs is to train the algorithm with normal data so that it can learn the dependency patterns of normal data, represented by the encoder and decoder weight matrices $W$ and bias vectors $b$ to compress and decompress the inputs. If the pattern of the test data does not match that of the normal data, utilizing the above-mentioned learnt non-linear matrices and vectors to encode and decode the corrupted data is likely to lead to a reconstruction error far greater than that of the `normal' data. In this way, the anomaly can be detected.

\subsection{Data flow in training and detection stages}

\begin{figure}[htp]
	\centering
	\includegraphics[scale=0.61]{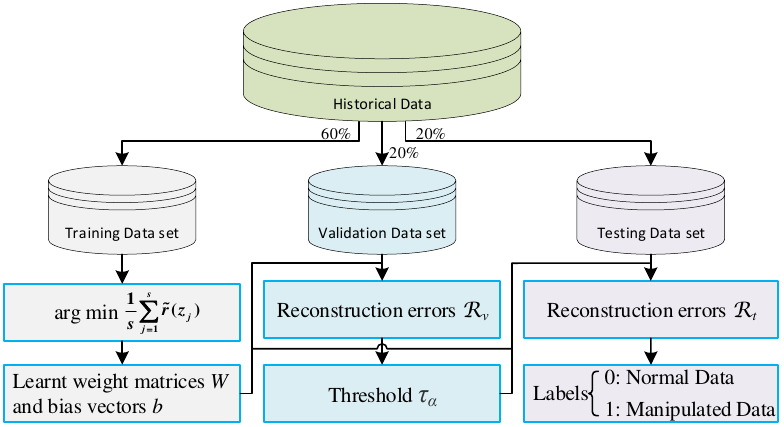}
	\caption{The data flow in the autoencoder neural network-based FDIA detector}
	\label{fig:flowchart}
\end{figure}

The data flow in our proposed FDIA detection network is depicted in Fig.~\ref{fig:flowchart}. The historical data are divided into training, validation and testing data sets with the proportion of 3:1:1. The weight matrices $W$ and bias vectors $b$ are updated in an iterative way with the goal of minimizing the mean value of all the reconstruction errors in the training data set as 
\begin{align}\label{opt:min_sum}
&\begin{array}{ccl}
& \min\limits_{W, \, b} & \Big\{ J := \frac{1}{S} \sum\limits_{j=1}^{S} \, {\tilde{r}}(z_j) \Big\} \, , \\
\end{array}
\end{align}
where $S$ denotes the total number of the observations used for training. After the convergence of $J$, the trained autoencoder network is utilized to encode and decode the validation data, resulting in the corresponding reconstruction errors $\mathcal{R}_v$, which are used to determine the threshold $\tau_{\alpha}$. Finally, the reconstruction errors $\mathcal{R}_t$ of the test data are compared with $\tau_{\alpha}$ to classify states into `normal' ($\tilde{r}(z) \le \tau_{\alpha}$) and `attack' (anomalous: $\tilde{r}(z) > \tau_{\alpha}$) data.

Because the autoencoder neural network is trained on normal data only, it can be considered an unsupervised one-class classifier. This overcomes the following challenges: (A) anomalous data is hard to obtain due to its rarity and confidentiality; (B) the variety of attack patterns makes it a time-consuming task to gather such patterns; (C) attack patterns are fast-evolving, so that detectors designed for known attack patterns may be unable to tackle new attack patterns.

\section{Case Study} 

 In this section, 
 experiments are conducted on the IEEE 118-bus system to evaluate the influence of hyperparameter selection on the training process and detection performance.

\subsection{Resource-constrained attack scenario }

We study an attack scenario from the perspective of an adversary that aims to interfere with the secure operation of the physical grid by manipulating the power flow measurements. By changing the apparent system state, the attacker can mislead the operator into taking costly or disruptive decisions. The attacker, in general, has limited resources while aiming to stay stealthy from the BDD. In light of this, we consider how many other measurements need to be attacked in coordination with the targeted power flow to avoid triggering alarms. This leads to a constrained optimization problem \cite{pan2018cyber}
\begin{align}\label{opt:sec_idx_o}
&\begin{array}{ccl}
& \min\limits_{a, \, c} & \lVert {a} \rVert_{0} \\
& \mbox{s.t.} & {a}= {Hc}, \ {a}_{i} = \mu, \\
& & {a}_{p} = 0, \ \forall p \in \mathcal{P},
\end{array}
\end{align}
where $\lVert {a} \rVert_{0}$ denotes the number of non-zero elements in attack vector $a$.  Here $\mu$ represents the value of injected false data on measurement $i$. We add the constraint that the measurements in the protected set $\mathcal{P}$ cannot be attacked. The computed optimal value of \eqref{opt:sec_idx_o} illustrates the minimum number of corrupted measurements in a stealthy attack against the measurement $i$. It is known that the above optimization program \eqref{opt:sec_idx_o} is non-convex and may be hard to solve in large problems. 
However, it can be expressed into a mixed integer linear program (MILP) which can be solved in an appropriate solver with acceptable computation time in an off-line manner.

The IEEE 118-bus system contains 99 loads, 54 generators and 186 transmission branches. For learning the normal operating conditions, the proposed mechanism is trained by the real data set which contains a total of 43,717 historical hourly loads from 32 European countries between 2013 and 2017 \cite{Muehlenpfordt2019}. 
These time series are used to generate 99 load point time series as described in \cite{Chenguang2020}. 
Interested readers can refer to this document for a detailed introduction to the method of generating normal operating conditions. The generated data set $\mathcal{T} \in \mathbb{R}^{43717 \times 339}$ was divided into a training set $\mathcal{T}_{r} \in \mathbb{R}^{26214 \times 339}$, a validation set $\mathcal{T}_v \in \mathbb{R}^{8743\times 339}$ and testing set $\mathcal{T}_t \in \mathbb{R}^{8760 \times 339}$.

\subsection{Hyperparameter tuning for training process }\label{subsec:tuning}
\begin{figure}[htp]
	\centering
	\includegraphics[scale=0.22]{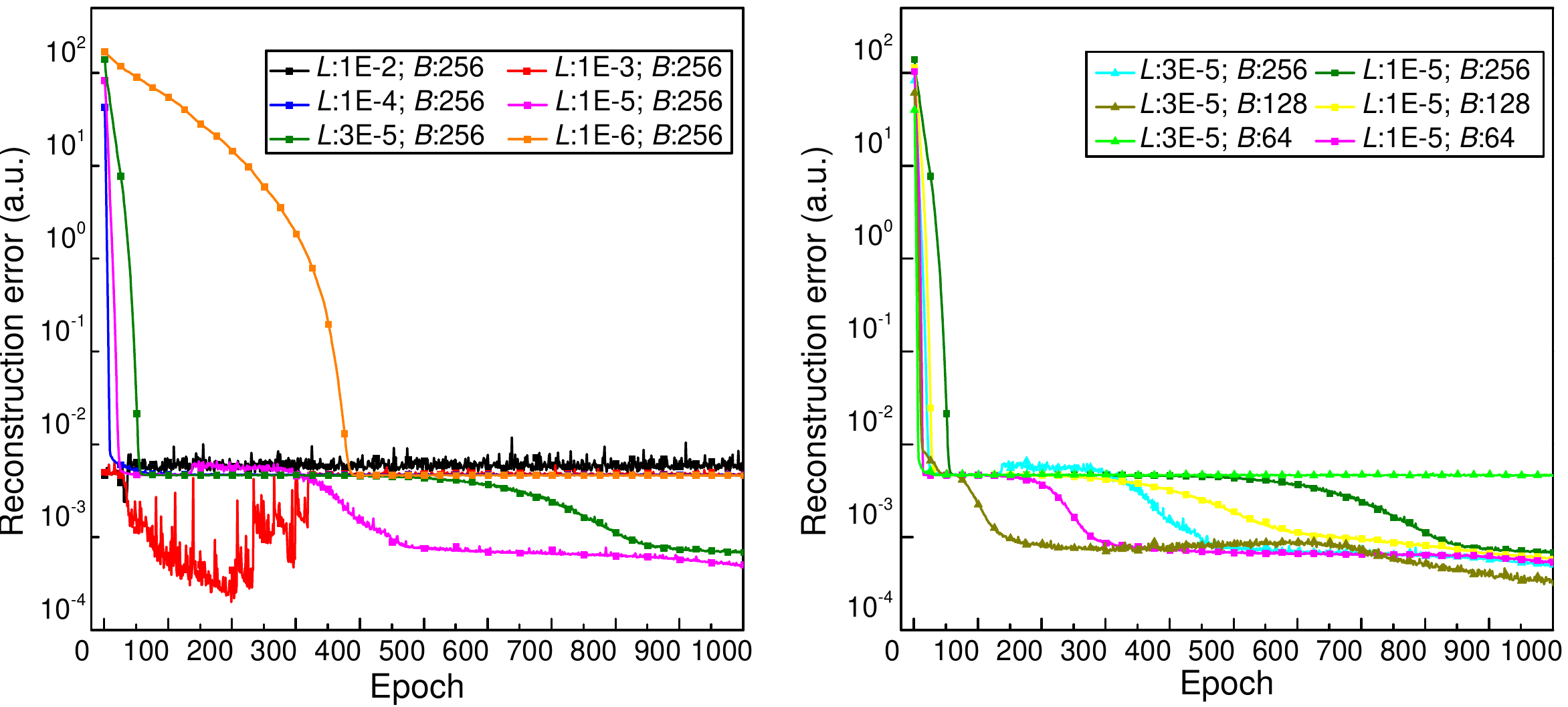}
	\caption{The relationship between the training epoch and the reconstruction error. $L$ stands for the learning rate and $B$ represents batch size.}
	\label{fig：Epoch}
\end{figure}

We tuned hyperparameters for the training process by using a grid search over learning rate ($10^{-2}$, $10^{-3}$, $10^{-4}$, $10^{-5}$) and batch size (64, 128, 256).  In addition, the encoder of autoencoder network is set to include 4 hidden layers, which are 339, 256, 128, and 64, respectively\cite{Chenguang2020}. The bottleneck layer has 32 nodes, and the decoder reconstructs the 32-dimensional data to a 339-dimensional output through 4 hidden layers with the same sizes as the encoder. In this paper, we utilize the sigmoid activation function from the second to penultimate hidden layer and the Adam Optimizer \cite{kingma2014adam} to iteratively optimize the value of weight matrices $W$ and bias vectors $b$. Training and testing of the autoencoder is conducted using \texttt{tensorflow} 2.1.0 on the Google Colab environment using the GPU option. The training performance under different parameters combination is shown in Fig.~\ref{fig：Epoch}.

We took batch size of 256 as an example to illustrate the trend of average reconstruction error $\tilde{r}(z)$ with the increase of training epoch. When the learning rate is set to be $10^{-2}$, $10^{-3}$ and $10^{-4}$, the mean value of reconstruction error converges to a high value or exhibits a fluctuation, which indicates high learning rates. However, when the learning rate is $10^{-6}$, it makes the convergence error of reconstruction error too slow. Therefore, $10^{-5}$ is selected as the appropriate value. Near $10^{-5}$, we looked for the appropriate learning rate at a higher resolution, and eventually, $10^{-5}$ and $3\times10^{-5}$, were set as candidates.

Then, we assign three different batch sizes (64, 128, 256) to the above alternative learning rates and compare the convergence performance. According to the results shown in Fig.~\ref{fig：Epoch}, in general, a high learning rate and small batch size result in a steeper reconstruction error convergence. However, if the two hyperparameters are set excessively for high convergence speed, the reconstruction error may unstably fluctuate during the decline (e.g. for learning rate  $3\times10^{-5}$ and batch size 128) or converge to a high value (e.g. for learning rate $3\times10^{-5}$ and batch size 64). In addition, a too-small batch size will increase iterations as well as the training time for running the same training epoch. Therefore, among the remaining two hyperparameter combinations (learning rate $10^{-5}$ and batch size 64;  
learning rate $3\times10^{-5}$ and batch size 256) that make the convergence fast and stable, we choose the latter for the training of our proposed detection mechanism.

\begin{figure}[htp]
	\centering
	\includegraphics[scale=0.35]{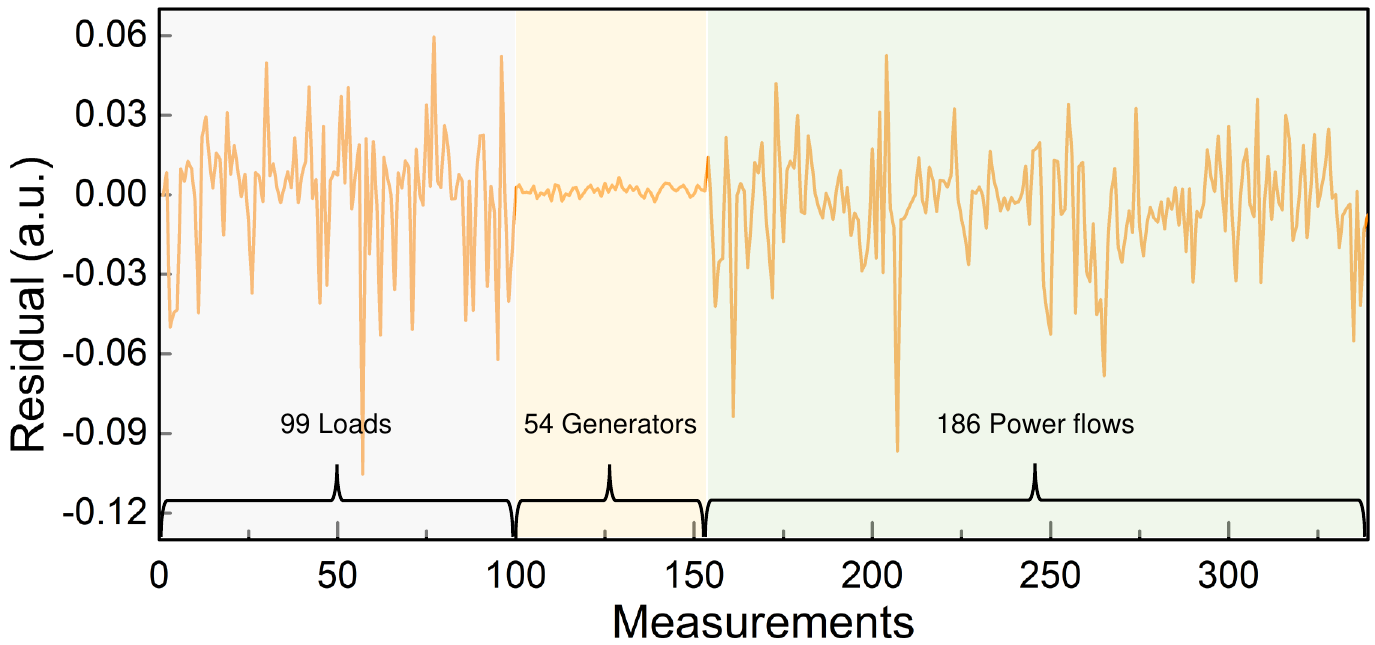}
	\caption{Residual of one observation}
	\label{fig:Reconstruction performance}
\end{figure}

Owing to the information loss that happens during encode and decode, there exists the residual between the measured data and its reconstruction value. The residuals of one normal observation which contains 339 measurements is depicted in Fig.~\ref{fig:Reconstruction performance}. Most residuals are in the range of -0.06 to 0.03. But it is worth noting that, to achieve a minimized mean value of reconstruction errors, a few measurements were ‘sacrificed', causing their residuals to be much higher than others. 

\subsection {Threshold selection strategy investigation} \label{subsec:scenario_2}

\begin{figure}[htp]
	\centering
	\includegraphics[scale=0.20]{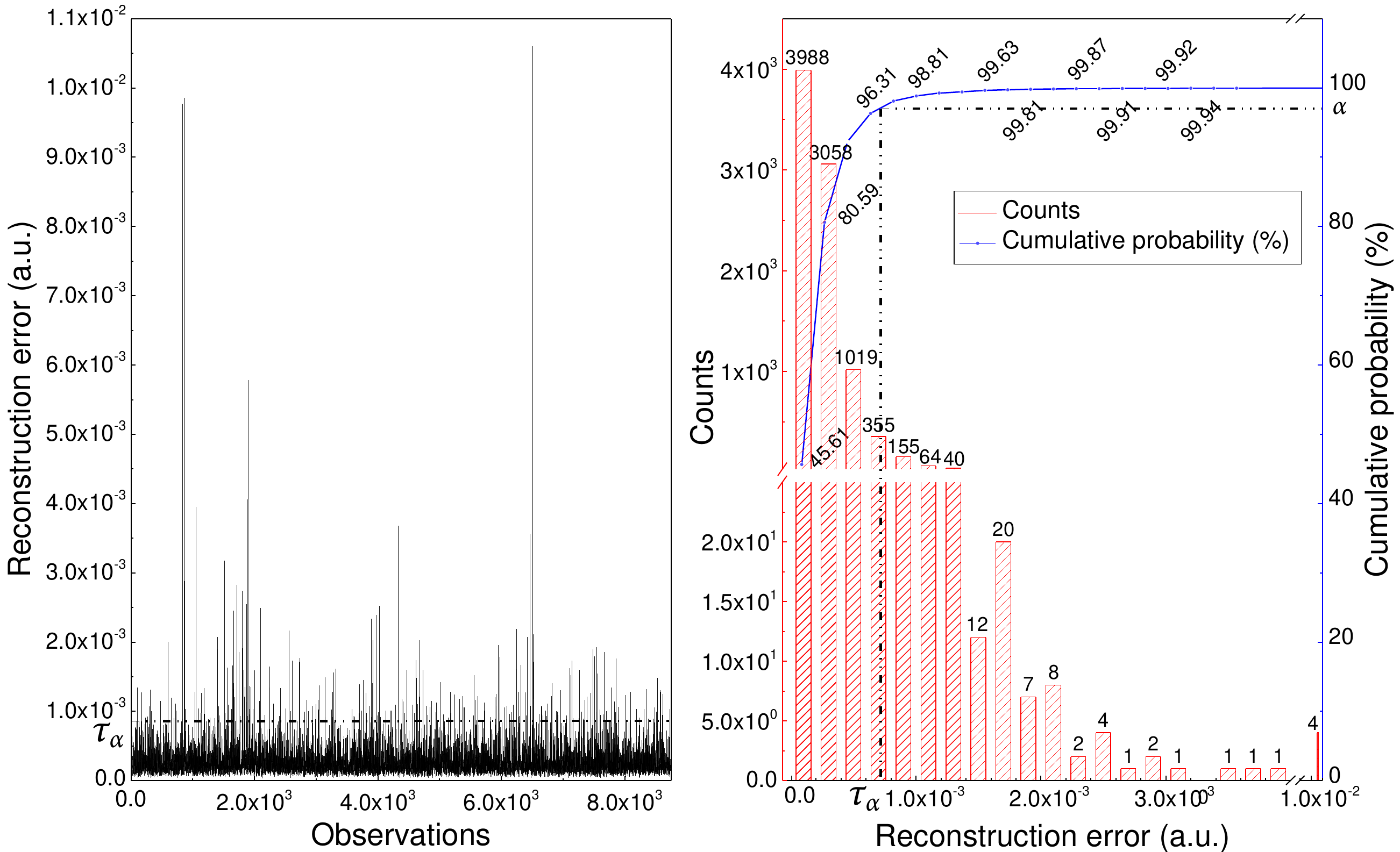}
	\caption{Reconstruction errors of the validation data set and the corresponding distribution}
	\label{fig:Reconstruction_errors_distribution}
\end{figure}

The autoencoder network was trained for 3000 epochs and the validation set confirmed an absence of overfitting.
The reconstruction errors of 8743 observations are calculated from their residuals by \eqref{eq:Re_error} and depicted in Fig.~\ref{fig:Reconstruction_errors_distribution}. After sorting ${R}_e$ in ascending order and observing the their distribution, a threshold $\tau_{\alpha}$ equal to the $\alpha^{th}$  percentile is chosen.

 In this study, we select the branch between bus 109 and 110 to launch power flow-targeted attacks. After solving the MILP of \eqref{opt:sec_idx_o}, the result shows the attacker needs to coordinately manipulate the measured power injection of bus 103, 109, 110 and the transmission line power flow from bus 103 to 110 at least. We launch 8760 attacks to manipulate hourly observations in the test data set $\mathcal{T}_t \in \mathbb{R}^{8760 \times 339}$ by decreasing the power flow from bus 109 to 110 by 10$\%$. Besides, we use the hourly uncorrupted normal operating data in $\mathcal{T}_t$ as a control group. Under power flow-targeted FDIAs, the influence of threshold selection on detection performance is shown in Table~\ref{tab:layer_percentile}. TP, FN, TN and FP denote true positive, false negative, true negative and false positive rate, respectively.

\renewcommand{\arraystretch}{1.5} %
\begin{table}[htp] 
                \caption{The influence  of  threshold  selection  on  power flow-targeted FDIA detection  performance.}
\begin{tabular}{p{0.02\textwidth}p{0.03\textwidth}p{0.08\textwidth}p{0.05\textwidth}p{0.05\textwidth}p{0.05\textwidth}p{0.05\textwidth}}
                  \hlineB{4} %
                              & $\alpha$ & $\tau_{\alpha}$ & TP & FN & TN  & FP    \\\hline
                              1& 96 & $7.67\times10^{-4}$ & $96.16\%$
 & 3.84\% & 92.05\% & 7.95\%  \\
                              2& 97 & $8.53\times10^{-4}$ & $95.62\%$
 & 4.38\% & 92.88\% & 7.12\%   \\
                              3& 98 & $ 9.89\times10^{-4}$ & $92.88\%$
 & 7.12\% & 94.25\% & 5.75\%    \\
                              4& 99 & $ 1.26\times10^{-3}$ & $91.78\%$
 & 8.22\% & 96.99\% & 3.01\%     \\ 
                              5& 99.5 & $ 1.67\times10^{-3}$ & $89.59\%$
 & 10.41\% & 98.08\% & 1.92\%     \\ 
                               6& 100 & $ 1.06\times10^{-2}$ & $67.67\%$
 & 32.33\% & 100.0\% & 0.00\%     \\ 
                \hlineB{4}
               \end{tabular}
               \label{tab:layer_percentile}
\end{table}

It can be observed that when $\alpha$ is increased from 95 to 100 percent, the false positive rate and true positive rate both decrease. In view of this,  $\alpha$ should be set to a sufficiently high value to decrease the false positive rate, but not so high that it comes at the cost of an excessive decrease in the true positive rate. From our experiment, it might be proper to choose an $\alpha$ near the inflection point where the cumulative distribution curve of reconstruction errors flattens out from the steep rise. This is consistent with the general practice in anomaly detection. In this case, the value of $\alpha$ is chosen as 99 to give consideration to both more hits (higher true positive rate) and fewer false alarms (lower false positive rate) as $91.78\%$ and $3.01\%$, respectively.

\subsection{Hyperparameter tuning and detection performance evaluation}\label{subsec:tuning_2}

In this experiment, we investigate the influence of hyperparameter selection, especially the depth and layer dimension of the proposed model on the FDIA detection performance.

\renewcommand{\arraystretch}{1.0} 

\begin{figure}[htp]
	\centering
	\includegraphics[scale=0.33]{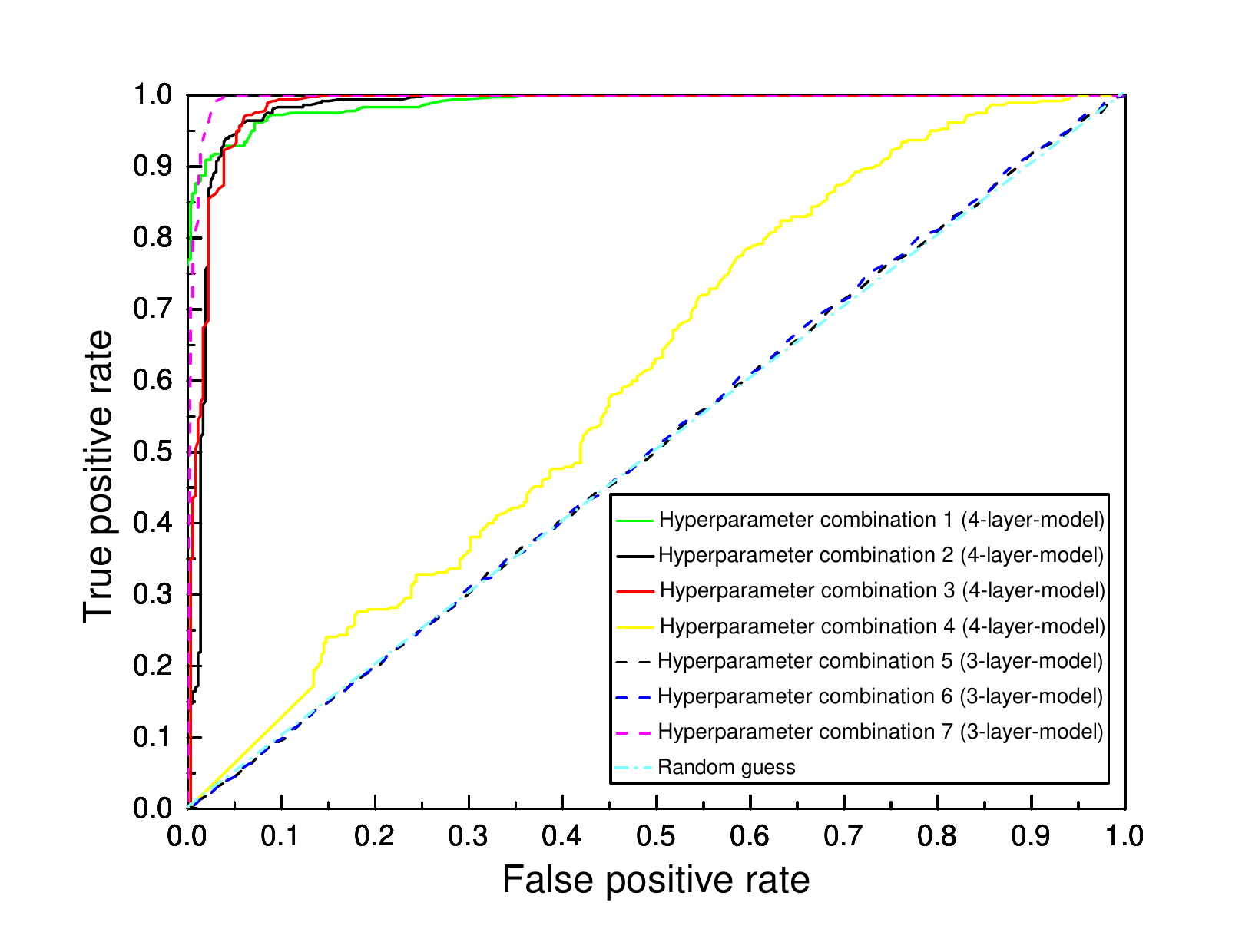}
	\caption{Receiver operating characteristic curves}
	\label{fig:ROC}
\end{figure}

\begin{table}[htp]  
               \caption{Hyperparameter combination and its FDIAs detection performance. $H$: Hidden layer, $B$: Bottleneck layer }
\begin{tabular}{p{0.02\textwidth}p{0.045\textwidth}p{0.045\textwidth}p{0.045\textwidth}p{0.045\textwidth}p{0.045\textwidth}p{0.08\textwidth}}
                  \hlineB{4} 
                              \multicolumn{7}{l}{\textbf{4-hidden-layer models}} \\ \hline
                              & $H_1$ & $H_2$ & $H_3$ & $H_4$ & $B$  & Avg. $\mathcal{R}_t$     \\
                              1  & 339 & 256 & 128 & 64  & 32 & $2.06\times10^{-4}$\\
                              2  & 339 & 256 & 128 & 64  & 24 & $2.30\times10^{-4}$\\
                              3  & 339 & 256 & 128 & 64  & 16 & $1.99\times10^{-4}$\\
                              4  & 339 & 256 & 128 & 64  & 8  & $4.31\times10^{-3}$\\ \hline
                              \multicolumn{7}{l}{\textbf{3-hidden-layer models}} \\ \hline
                              & $H_1$ & $H_2$ & $H_3$ &  & $B$  & Avg. $\mathcal{R}_t$      \\
                              5  & 339 & 128 & 64  &     & 32 & $4.68\times10^{-3}$  \\
                              6  & 339 & 256 & 64  &     & 32 & $4.67\times10^{-3}$  \\
                              7  & 339 & 256 & 128  &     & 32 & $1.90\times10^{-4}$ \\
                              \hlineB{4}
               \end{tabular}
               \label{tab:layer_3_4}
\end{table}
 We consider 3 and 4-layer models with 7 different dimension configuration combinations as shown in Table~\ref{tab:layer_3_4}. In 4-layer-models, we only change the dimension of the bottleneck layer. For the 3-layer-models, the difference exists in the dimension combination of the second and third hidden layer. Other training hyperparameters remain the same as in Section~\ref{subsec:tuning} and the attack target remains unchanged from the previous experiments in Section~\ref{subsec:scenario_2}. The result is shown in Fig.~\ref{fig:ROC} as receiver operating characteristic curves (ROC) to compare the detection sensitivity (true positive rate) and specificity (false positive rate) under different model configurations. 
 
 In 4-layer-models, as the dimension of the bottleneck layer decreases from 32 to 16, the models still demonstrate satisfactory detection performance overall. However, if the model is over compressed into the latent space as a 8-dimensional bottleneck layer, it will result in the excessive loss of  information during the encoding/decoding process and thus interfere with the detection accuracy.  As for the 3-layer-models, the reduction of layers (hyperparameter combinations 5 and 6) may lead to the increase of the reconstruction error which is shown in the last column of Table~\ref{tab:layer_3_4} and gives rise to the decline of detection sensitivity and specificity. However, the model with hyperparameter combination 7 still denotes comparable detection probability and false-alarm probability as 4-layer-models. This indicates that to set the dimensions of each layer properly, in particular, to reduce the dimensional gap between layers helps to enhance the model's reconstruction and detection capabilities.

\section{Conclusion}

In this paper, we describe an autoencoder neural network which uses normal operating data only, and tune the  hyperparameters to detect power flow-targeted FDIAs. The main contribution is that we investigate the influence of the hyperparameter selection on the training process and the FDIA detection performance, and put forward preliminary  hyperparameter selection and tuning strategies. The experimental results demonstrate that, if it is configured properly, the mechanism is able to demonstrate satisfactory learning efficiency and detection effectiveness. In future work, we aim to investigate automated and computationally efficient hyperparameter tuning strategies, and the impact of the choice of reconstruction error metrics. We will increase the complexity of case studies to further investigate the robustness of the proposed approach. 

%


\bibliographystyle{IEEEtran}
\bibliography{literature}

\end{document}